# Approximating percentage of academic traffic in the World Wide Web and rankings of countries based on academic traffic

Bahram Kalhor [1], Mohammad Reza Ghane[2], Alireza Nikravanshalmani[1]

[1]Department of Computer, College of Mechatronic, Karaj Branch, Islamic Azad University, Alborz, IRAN~~Department of Computer Science, Tehran Science~~

[2]Regional Information Center for Science and Technology (RICeST), Shiraz, IRAN

Kalhor_bahram@yahoo.com; ghane@ricest.ac.ir; nikravan@kiau.ac.ir

**00989123446896 (Kalhor)**

# Approximating percentage of academic traffic in the World Wide Web and rankings of countries based on academic traffic

**Abstract**

The paper introduces a novel mechanism for approximating traffic of the academic sites (universities and research institutes) in the World Wide Web based on Alexa's rankings. Firstly we introduce and discuss new method for calculating score (weight) of each site based on its Alexa's rank. Secondly we calculate percentage of academic traffic in the World Wide Web. Thirdly we introduce and discuss two new rankings of countries based on academic traffic. Finally we discuss about three indicators and effects of them in traffic of the academic sites. Results indicate that the methodology can be useful for approximating traffic of the academic sites and producing rankings of countries in practice.

1. **Introduction**

   Traffic of a site is amount of visitors and visits a site receives. Percentage of academic traffic in the World Wide Web has not yet been calculated and there are no rankings of countries based on academic traffic in the World Wide Web. The primary method for calculating traffic of the sites is counting hits of the visitors. Unfortunately, most of the site explorers don't provide the number of hits. Although, Alexa is the most famous site which provides rank of sites' traffic, Alexa provides global rank of the sites without the number of the hits. This research introduces new indicator that estimate score of the sites' traffic based on Alexa's rank and approximate traffic of the academic sites in the World Wide Web based on Alexa's rankings.
   Rankings of universities are a quantitative style of universities performance evaluation (Huang, 2011). Higher education institutions are using these rankings to show their educational, research or business excellence performance (Isidro F. Aguillo et al., 2010). Higher education systems in competitive environments generally present top universities (GiovanniAbramo et al., 2011). There is an increasing interest in rankings of universities (Isidro F. Aguillo et al., 2010). For this purpose a growing use of impact metrics in the evaluation of scholars, journals and academic institutions and even countries has been occurred (JasleenKaur, FilippoRadicchi, FilippoMenczer, 2013). Different indicators which have been introduced based on hyperlinks, citations, scholars and papers are investigated for ranking universities. Scientific publication has moved to the web, and novel approaches to scholarly communication (Blaise Cronin, 2011). There are several large scale university ranking programs (Chi-ShiouLin, Mu-HsuanHuang, Dar-ZenChen, 2012).
   Traffic of a site is one of the most important indicators, which can be used to evaluate sites' effectiveness. A few number of research have been published which use the web traffic as an indicator of site's ranking.

Numbers of works have used web traffic for this purpose (Ortega & Aguillo, 2009; Vaughan & Yang, 2013; Wolk & Teysohn, 2007; Plaza, 2009; Turner, 2010). The distribution of the number of hits and the spent time of web sessions are characterized in (Ortega & Aguillo, 2009). In (Vaughan & Yang, 2013) the significant correlation is investigated between web traffic and academic quality. They have gained the sits' ranking from three sources: Alexa, Compete and Google trend, and found that there is a significant correlation between web traffic and academic quality. They also concluded that Alexa is more reliable than two other sites. Different factors that influence the website traffic in the paid content market, have been analyzed by (Wolk & Teysohn, 2007). Google Analytics have been used by (Plaza, 2009) as traffic data source. Although Google Trend have been used by (Spencer, 2011), Google Trend has received little attention in researches. Alexa have been used by (Callaway, 2011) for comparing bank performance. Relation between Web site's traffic of banks and the performance of the banks is shown with Callaway. Traffic of the sites is used as an indicator to compare bank performance

4icu.org provides the global rankings of universities, which rank universities in whole world (http://www.4icu.org/about/). Regional Information Center for Science and Technology (en.ricest.ac.ir) in IRAN provides the local rank of IRAN's universities. In both ranking schemes, web traffic is used as an important indicator. To our best knowledge, there hasn't been any study for ranking countries based on universities' traffic.

For calculating the academic traffic percentage in the world, having the universities' rank is the key factor in our work. Our main contribution in this study is to use the traffic rank of universities and propose a new schema for ranking countries based on academic traffic for the first time.

2. **Data and methods**

For approximating traffics of countries and estimating percentage of academic traffic in the world, rankings of traffics for more than 21000 universities and research centers have been used in this study.

*2.1. Data source*

Alexa provides rankings of sites based on visitor's hits. Alexa's traffic estimates are based on a diverse sample of millions of worldwide internet users (http://www.alexa.com/company). Traffic data of Alexa are gathered from computers which the Alexa's toolbar is installed on them. Alexa provides two traffic rankings. The first rank is site's rank according to visitors of the country, and the other is global rank which is site's rank according to visitors around the world. Alexa tracks over 30 million websites. Sorting is based on the 3 month Alexa traffic rank. In this study we have used global rank of sites. We assigned Alexa traffic rank equal to 30,000,000 for universities which don't have Alexa global rank. Alexa publishes the Top 500 Web sites at http://www.alexa.com/topsites/global.

*2.2. Selecting countries and universities*

In this study our database contains 21,485 universities, which is the same as the universities announced in to web rankings of world universities (WR, http://webometrics.info). WR is a web based rankings of universities which contains broad list of universities from all over the world (January 2014, 21,451 universities).

All countries which have at least one university in our database are considered in this study. 21,485 sites of universities and research centers have been covered by 197 countries and 1 international category (5 universities).

These numbers seem too large in some cases. For instance, Finland has 16 universities and 29 professional university colleges

## 2.3. Weight of universities

Alexa only provides the rankings of sites. It doesn't give the computed score of each site that has been used for their ranking method. Weighting universities is utilized to approximate the number of hits. It is also used to rank of countries and to compute academic traffic to this end. $W_u$ is proposed as the new indicator which is used to detect the weight of each university.

$$W_u = 1 - \frac{R_u}{M}$$

where $W_u$ is weight of university, $R_u$ is global rank of university in Alexa and M is maximum rank of sites according to Alexa (M=30,000,000),

Min $W_u = 0$ and

Max $W_u = 1-(1/30,000,000) = 0.9999999666667$.

## 2.4. Weight of countries

The proposed new indicator $W_u$ is then used to calculate the weight of each country. The new formula for this purpose is proposed as follow:

$$W_c = \sum_{i=1}^{n} W_{u_i}$$

where $W_c$ is country's weight, n is the number of universities in the country and $W_{u_i}$ is the weight of i th university.

## 2.5. Weight of academic traffic

Academic traffic is computed according to weight of all universities.

$$W_{at} = \sum_{i=1}^{k} W_{u_i}$$

Where $W_a$ is weight of academic traffic, k is number of universities in the world (21,485 in this study, January 2014), $W_{u_i}$ is weight of university number (i).

## 2.6. $P_{at}$: Percentage of academic traffic

Finally, we introduce $P_{at}$ as a new indicator for calculating academic traffic in the world, based on universities rank in Alexa and maximum rank of 30 million sites in Alexa. The proposed formula is as follows:

$$P_{at} = \frac{W_{at}}{\sum_{j=1}^{M} \frac{1}{j}} * 100$$

where

M is the total number of sites according to Alexa,

$W_{at}$ is the total weights of universities,

$\sum_{j=1}^{M} \frac{1}{j}$ is the total weights of sites and

$P_{at}$ is the percentage of academic traffic.

2.7. Average weight of countries

New formula has been proposed for comparing countries based on number of universities and academic's weight of countries. Average rank of each country has been calculated by dividing the weight of country to number of universities. Table (4) shows the result of average weight of countries (average traffic rank) for all countries which have at least 100 universities.

$$A_{wc} = \frac{\sum_{i=1}^{n} W_{u_i}}{n}$$

where

n is the total number of university in each country,

$W_{ui}$ is the weight of university,

$A_{wc}$ is the average weight of country.

3. Results

Table (1) shows the top 60 countries with highest number of active universities. There are some differences between real universities count in each country with data of table (1) which has been collected from webometrics.info. For instance, Finland has 16 universities and 29 professional university colleges. In this table United States of America, Brazil, India, China and Russian Federation are at the top of the list.

Google has been used for creating map of the world with separated countries. Countries' data which are collected at January 2014 have been saved in private MS Access database. Each country's name has a unique two character which is standard in the world. Traffic data of the all countries and the two characters code have been sent to Google for creating the color map of the countries. Figure (1) has been created depend on real data of the all countries based on number of universities in each country.

**Table 1:**

Top 60 countries with highest number of universities

| Rank | Country Name | Number of Universities | Rank | Country Name | Number of Universities | Rank | Country Name | Number of Universities |
|---|---|---|---|---|---|---|---|---|
| 1 | United States of America | 3344 | 21 | Italy | 225 | 41 | Malaysia | 82 |
| 2 | Brazil | 1834 | 22 | Thailand | 183 | 42 | Hungary | 82 |
| 3 | India | 1743 | 23 | Turkey | 170 | 43 | Austria | 77 |
| 4 | China | 1252 | 24 | Taiwan | 170 | 44 | Georgia | 77 |
| 5 | Russian Federation | 1088 | 25 | Netherlands | 156 | 45 | Bosnia and Herzegovina | 75 |
| 6 | Mexico | 962 | 26 | Nigeria | 144 | 46 | Algeria | 74 |
| 7 | Japan | 861 | 27 | Vietnam | 124 | 47 | Ecuador | 72 |
| 8 | France | 635 | 28 | Kazakstan | 120 | 48 | Venezuela | 71 |
| 9 | Iran (Islamic Republic of Iran) | 605 | 29 | Portugal | 118 | 49 | Norway | 67 |
| 10 | Poland | 475 | 30 | Argentina | 117 | 50 | Greece | 67 |
| 11 | Germany | 425 | 31 | Switzerland | 113 | 51 | Saudi Arabia | 62 |
| 12 | Republic Of Korea | 419 | 32 | Romania | 111 | 52 | Egypt | 60 |
| 13 | Indonesia | 373 | 33 | Bangladesh | 107 | 53 | Costa Rica | 59 |
| 14 | Pakistan | 344 | 34 | Morocco | 105 | 54 | Bulgaria | 59 |
| 15 | Ukraine | 336 | 35 | Australia | 104 | 55 | Uzbekistan | 58 |
| 16 | United Kingdom | 330 | 36 | Belgium | 99 | 56 | Latvia | 58 |
| 17 | Philippines | 307 | 37 | Denmark | 98 | 57 | Finland | 57 |
| 18 | Colombia | 306 | 38 | Peru | 92 | 58 | Belarus | 56 |
| 19 | Canada | 265 | 39 | Czech Republic | 85 | 59 | Tunisia | 55 |
| 20 | Spain | 248 | 40 | Chile | 85 | 60 | Iraq | 55 |

**Figure 1:**

Countries are colored based on number of universities in each country

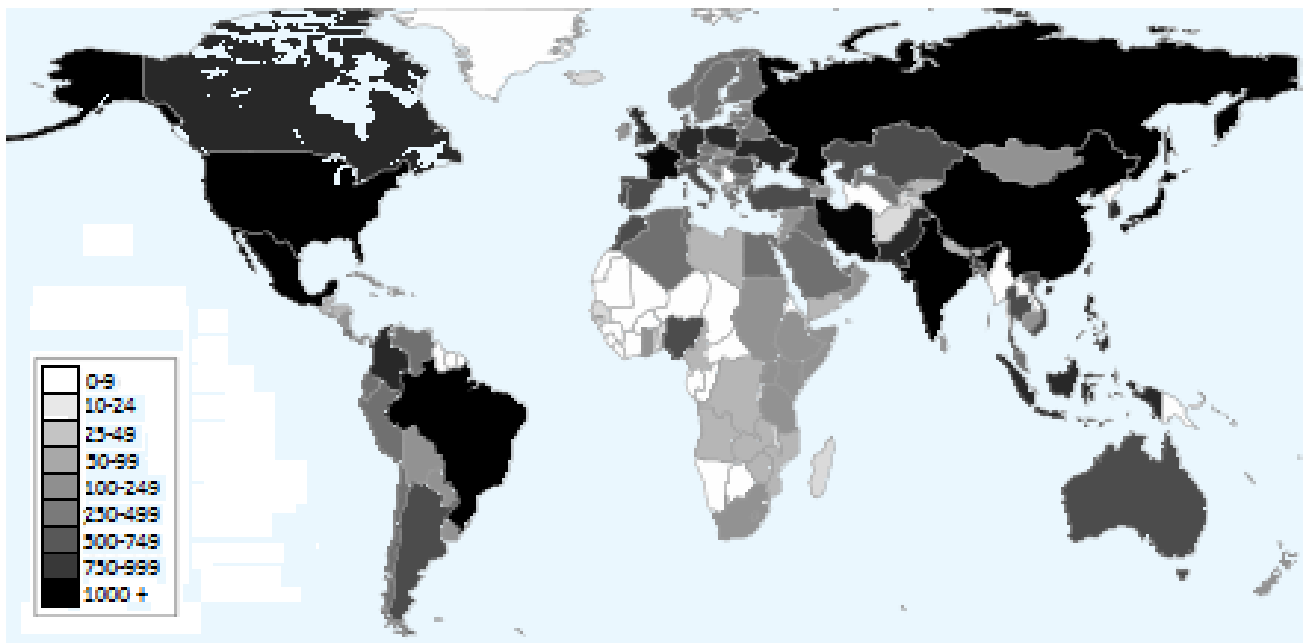

Table (2) shows the top 80 universities and their associated weights which is taken in January 2014.

**Table 2:**
Weight of top 80 universities

| Rank | University Name | $W_u$ | Rank | University Name | $W_u$ |
|---|---|---|---|---|---|
| 1 | Singapore-MIT Alliance for Research and Technology | 0.000864 | 41 | Yale University | 0.000214 |
| 2 | Harvard-MIT Division of Health Sciences and Tecnology | 0.000858 | 42 | Carnegie Mellon University | 0.000207 |
| 3 | Massachusetts Institute of Technology | 0.000852 | 43 | Centro Universitario Estadual da Zona Oeste | 0.000197 |
| 4 | Stanford University | 0.000714 | 44 | University of Florida | 0.000197 |
| 5 | Harvard University | 0.000644 | 45 | University of Oxford | 0.000189 |
| 6 | Universidad Nacional Mexico | 0.000509 | 46 | University of Applied Science and Technology Tehran | 0.000178 |
| 7 | University of California Berkeley | 0.000463 | 47 | University of Toronto | 0.000175 |
| 8 | Pennsylvania State University | 0.000419 | 48 | Victoria University in the University of Toronto | 0.000174 |
| 9 | Columbia University New York | 0.000373 | 49 | Trinity College in the University of Toronto | 0.000174 |
| 10 | Cornell University | 0.000346 | 50 | University of Southern California | 0.000171 |
| 11 | Weill Medical College Cornell University | 0.000341 | 51 | Harvard University Harvard Business School | 0.000163 |
| 12 | Weill Cornell Medical College in Qatar | 0.000340 | 52 | Ohio State University | 0.000161 |
| 13 | University of Texas Austin | 0.000314 | 53 | University of British Columbia | 0.000161 |
| 14 | University of Michigan | 0.000302 | 54 | University of California Davis | 0.000156 |
| 15 | New York University | 0.000300 | 55 | Payam Noor University Kabodrahang | 0.000151 |
| 16 | University of Michigan Dearborn | 0.000298 | 56 | Payam Noor University | 0.000150 |
| 17 | University of Wisconsin Madison | 0.000258 | 57 | University of Cambridge | 0.000149 |
| 18 | University of Pennsylvania | 0.000253 | 58 | University of California San Diego | 0.000148 |
| 19 | University of Minnesota | 0.000251 | 59 | Rutgers University | 0.000148 |
| 20 | University of Washington | 0.000250 | 60 | University of Phoenix | 0.000147 |
| 21 | University of Minnesota Duluth | 0.000249 | 61 | Universidade Paulo USP | 0.000147 |
| 22 | University of Minnesota Morris | 0.000247 | 62 | Michigan State University | 0.000147 |
| 23 | University of Minnesota Crookston | 0.000247 | 63 | Academy of State Fire Service | 0.000146 |
| 24 | University of Minnesota, Rochester | 0.000246 | 64 | Rutgers University Camden | 0.000144 |
| 25 | University of Illinois Urbana Champaign | 0.000246 | 65 | University of Tehran | 0.000143 |
| 26 | University of California Los Angeles UCLA | 0.000244 | 66 | Palawan State University | 0.000139 |
| 27 | Purdue University | 0.000243 | 67 | Universidad Santo Tom | 0.000138 |
| 28 | Princeton University | 0.000243 | 68 | Escuela de Arquitectura de Chihuahua | 0.000138 |
| 29 | Moscow Regional Social and Economic Institute | 0.000243 | 69 | University of Maryland | 0.000137 |
| 30 | CUNY Medgar Evers College | 0.000217 | 70 | University of North Carolina Chapel Hill | 0.000135 |
| 31 | CUNY John Jay College of Criminal Justice | 0.000217 | 71 | North Carolina State University | 0.000133 |
| 32 | City University of New York | 0.000217 | 72 | Arizona State University | 0.000133 |
| 33 | CUNY York College | 0.000217 | 73 | College of Law Latvia | 0.000132 |
| 34 | CUNY New York City College of Technology | 0.000217 | 74 | Boston University | 0.000132 |
| 35 | CUNY Queens College | 0.000217 | 75 | Escola de Governo Professor Paulo Neves de Carvalho | 0.000132 |
| 36 | CUNY Brooklyn College | 0.000217 | 76 | Helena Antipoff FHA | 0.000132 |
| 37 | CUNY Hunter College | 0.000216 | 77 | Trabalho de Minas Gerais UTRAMIG | 0.000132 |
| 38 | City College of New York CUNY | 0.000216 | 78 | University of Anatolia | 0.000130 |
| 39 | CUNY College of Staten Island | 0.000216 | 79 | Duke University | 0.000129 |
| 40 | CUNY Baruch College | 0.000216 | 80 | University of Arizona | 0.000127 |

Table (3) shows top 60 countries with highest calculated weights. United State of America has gained 38.6 percent of academic traffic of the world.

Although, countries with more universities have higher chance to have better rank, but, having more universities doesn't guarantee to take the better rank, for example Russian Federation which is the 5'th country in table (1), sits in the 11'th position in table (3).

Figure (2) shows colored countries which are based on weights of each country. In figure (2) the color of each country are selected based on weight of academic traffic in table (3).

**Table ٣:**
Top 60 countries with highest weight of academic traffic

| Rank | Country Name | Number of Universities | Weight of Country ($W_c$) | Rank | Country Name | Number of Universities | Weight of Country ($W_c$) |
|---|---|---|---|---|---|---|---|
| 1 | United States of America | 3344 | 0.03398612 | 31 | Chile | 85 | 0.00043510 |
| 2 | India | 1743 | 0.00443464 | 32 | Republic Of Korea | 419 | 0.00040633 |
| 3 | Brazil | 1834 | 0.00405049 | 33 | Malaysia | 82 | 0.00040565 |
| 4 | China | 1252 | 0.00342308 | 34 | Greece | 67 | 0.00038486 |
| 5 | Iran (Islamic Republic of Iran) | 605 | 0.00336076 | 35 | Belgium | 99 | 0.00037599 |
| 6 | United Kingdom | 330 | 0.00294720 | 36 | Austria | 77 | 0.00034636 |
| 7 | Canada | 265 | 0.00277408 | 37 | Peru | 92 | 0.00033916 |
| 8 | Germany | 425 | 0.00265298 | 38 | Finland | 57 | 0.00032652 |
| 9 | Spain | 248 | 0.00206200 | 39 | Singapore | 34 | 0.00032567 |
| 10 | France | 635 | 0.00205915 | 40 | Czech Republic | 85 | 0.00031367 |
| 11 | Russian Federation | 1088 | 0.00195689 | 41 | South Africa | 26 | 0.00030696 |
| 12 | Japan | 861 | 0.00192931 | 42 | Portugal | 118 | 0.00029652 |
| 13 | Mexico | 962 | 0.00177423 | 43 | Norway | 67 | 0.00029338 |
| 14 | Australia | 104 | 0.00123761 | 44 | Vietnam | 124 | 0.00029068 |
| 15 | Italy | 225 | 0.00122493 | 45 | Philippines | 307 | 0.00025494 |
| 16 | Indonesia | 373 | 0.00113263 | 46 | Hong Kong | 24 | 0.00024370 |
| 17 | Turkey | 170 | 0.00105685 | 47 | Bangladesh | 107 | 0.00024084 |
| 18 | Taiwan | 170 | 0.00091697 | 48 | Kazakstan | 120 | 0.00022972 |
| 19 | international | 5 | 0.00087933 | 49 | Nigeria | 144 | 0.00022295 |
| 20 | Poland | 475 | 0.00081460 | 50 | Ukraine | 336 | 0.00021627 |
| 21 | Saudi Arabia | 62 | 0.00073428 | 51 | Denmark | 98 | 0.00021607 |
| 22 | Colombia | 306 | 0.00060642 | 52 | Egypt | 60 | 0.00020863 |
| 23 | Cuba | 30 | 0.00058395 | 53 | Ireland | 49 | 0.00020186 |
| 24 | Netherlands | 156 | 0.00056501 | 54 | Venezuela | 71 | 0.00019839 |
| 25 | Pakistan | 344 | 0.00055320 | 55 | Latvia | 58 | 0.00016995 |
| 26 | Qatar | 7 | 0.00053197 | 56 | Hungary | 82 | 0.00016972 |
| 27 | Thailand | 183 | 0.00046288 | 57 | Bolivia | 48 | 0.00016781 |
| 28 | Switzerland | 113 | 0.00045083 | 58 | New Zealand | 42 | 0.00016323 |
| 29 | Sweden | 53 | 0.00044227 | 59 | Romania | 111 | 0.00016062 |
| 30 | Argentina | 117 | 0.00044173 | 60 | Belarus | 56 | 0.00013981 |

**Figure 2:**
Countries are colored based on traffic's weight of each country

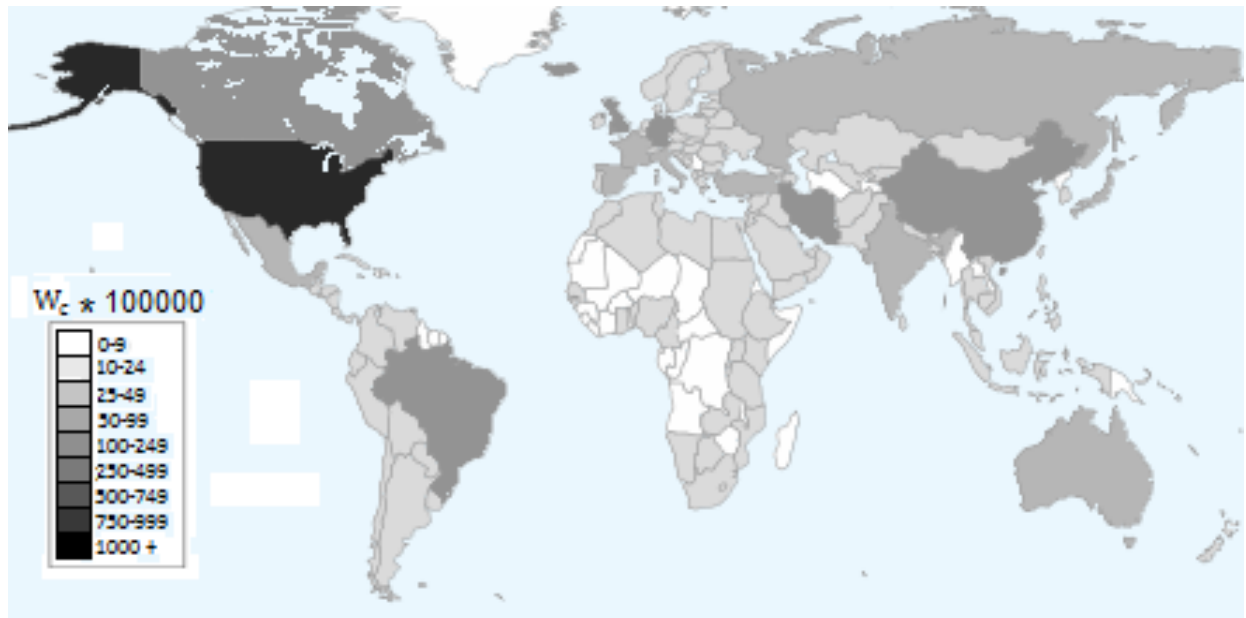

**Table 4:**
Average weight of all universities in each country

| Country Name | Country's Weight | Number of Universities | Average Weight of Country $(A_{wc}) * 10^6$ |
|---|---|---|---|
| Australia | 0123761 | 104 | 11.90 |
| Canada | 0277408 | 265 | 10.47 |
| United States of America | 0.03398612 | 3344 | 10.16 |
| United Kingdom | 0294720 | 330 | 8.93 |
| Spain | 0206200 | 248 | 8.31 |
| Germany | 0265298 | 425 | 6.24 |
| Turkey | 0105685 | 170 | 6.22 |
| Iran (Islamic Republic of Iran) | 0336076 | 605 | 5.55 |
| Italy | 0122493 | 225 | 5.44 |
| Taiwan | 0091697 | 170 | 5.39 |
| Switzerland | 0045083 | 113 | 3.99 |
| Argentina | 0044173 | 117 | 3.78 |
| Netherlands | 0056501 | 156 | 3.62 |
| France | 0205915 | 635 | 3.24 |
| Indonesia | 0113263 | 373 | 3.04 |
| China | 0342308 | 1252 | 2.73 |
| India | 0443464 | 1743 | 2.54 |
| Thailand | 0046288 | 183 | 2.53 |
| Portugal | 0029652 | 118 | 2.51 |
| Vietnam | 0029068 | 124 | 2.34 |
| Bangladesh | 0024084 | 107 | 2.25 |
| Japan | 0192931 | 861 | 2.24 |
| Brazil | 0405049 | 1834 | 2.21 |
| Colombia | 0060642 | 306 | 1.98 |
| Kazakstan | 0022972 | 120 | 1.91 |
| Mexico | 0177423 | 962 | 1.84 |
| Russian Federation | 0195689 | 1088 | 1.80 |
| Poland | 0081460 | 475 | 1.71 |

| Country Name | Country's Weight | Number of Universities | Average Weight of Country $(A_{wc}) * 10^6$ |
|---|---|---|---|
| Pakistan | 0055320 | 344 | 1.61 |
| Nigeria | 0022295 | 144 | 1.55 |
| Romania | 0016062 | 111 | 1.45 |
| Republic Of Korea | 0040633 | 419 | 0.97 |
| Philippines | 0025494 | 307 | 0.83 |
| Ukraine | 0021627 | 336 | 0.64 |
| Morocco | 0006231 | 105 | 0.59 |

By substituting the real values in above formula the total weights of universities' sites equal to 0.0890988, the total weights of all sites is equal to 17.7939 and $P_{at}$ is equal to 0.50072 %.

$P_{at}$=0.5% shows academic traffic of the 21,485 academic sites in the 30 million sites. The results show that about 0.5 percent of all traffics of the World Wide Web belong to academic traffic. One hit of each 200 hits in the internet, belongs to academic sites. 21,485 sites are 0.073 percent of all Alexa's sites (by dividing 21,485 to 30 millions) but 0.5 percent of traffics belong to academic web sites. By dividing 0.5 to 0.073 we will reach to 6.85, this means average of hitting academic web sites are 6.85 times of average hits per all sites around the world.

United States of America, India, Brazil, China and Iran (Islamic Republic of Iran) are at the top of the list of countries which use most academic traffic of the world. 38.6 percent of academic traffic belongs to United State of America.

Australia, Canada, United States of America, United Kingdom and Spain are 5 countries with highest average weight of countries' academic traffic.

4. Discussion

Rankings of counties based on academic traffic have been investigated in this study. As mentioned in the previous sections, academic traffic rank of countries has been calculated based on universities traffic rank. In this case, countries with higher number of universities have been more chance to gain more weight. The selected approach hides important information. Two university sites close to each other regarding ranks might be far from each other with respect to number of hits. Conversely, two university sites far from each other regarding ranks might be close to each other with respect to number of hits. For overcoming this problem we have used highest number of universities. In this case, differences between ranks are at minimum quantity then two university sites close to each other regarding ranks are close to each other with respect to number of hits.

More universities don't guarantee to gain better rank, for example Russian Federation which is the 5'th country in table (1), sits in the 11'th position in table (3). In the other hand some countries with lower number of universities, sit in higher rank in table (3), for example Iran which is the 9'th country in table (1), sits in the 5'th position in table (3).

Now, the main question is: What are the most important indicators for countries to take better academic traffic rank? The number of population, number of universities and performance of sites are some candidates which have been more considered to study in this section.

List of countries and their populations are presented in table (5) and map of the world population is presented in right part of figure (3) based on World Bank, which shows a relation between number of countries' population and number of universities in each country.

**Table 5:**
List of countries by population
(http://data.worldbank.org/indicator/SP.POP.TOTL?order=wbapi_data_value_2013+wbapi_data_value+wbapi_data_value-last&sort=desc)

|   | Country name | 2009 | 2010 | 2011 | 2012 | 2013 |
|---|---|---|---|---|---|---|
| 1 | China | 1,331,260,000 | 1,337,705,000 | 1,344,130,000 | 1,350,695,000 | 1,357,380,000 |
| 2 | India | 1,190,138,069 | 1,205,624,648 | 1,221,156,319 | 1,236,686,732 | 1,252,139,596 |
| 3 | United States | 306,771,529 | 309,326,295 | 311,582,564 | 313,873,685 | 316,128,839 |
| 4 | Indonesia | 237,486,894 | 240,676,485 | 243,801,639 | 246,864,191 | 249,865,631 |
| 5 | Brazil | 193,490,922 | 195,210,154 | 196,935,134 | 198,656,019 | 200,361,925 |
| 6 | Pakistan | 170,093,999 | 173,149,306 | 176,166,353 | 179,160,111 | 182,142,594 |
| 7 | Nigeria | 155,381,020 | 159,707,780 | 164,192,925 | 168,833,776 | 173,615,345 |
| 8 | Bangladesh | 149,503,100 | 151,125,475 | 152,862,431 | 154,695,368 | 156,594,962 |
| 9 | Russian Federation | 141,909,244 | 142,385,523 | 142,956,460 | 143,178,000 | 143,499,861 |
| 10 | Japan | 127,557,958 | 127,450,459 | 127,817,277 | 127,561,489 | 127,338,621 |
| 11 | Mexico | 116,422,752 | 117,886,404 | 119,361,233 | 120,847,477 | 122,332,399 |
| 12 | Philippines | 91,886,400 | 93,444,322 | 95,053,437 | 96,706,764 | 98,393,574 |
| 13 | Ethiopia | 84,838,032 | 87,095,281 | 89,393,063 | 91,728,849 | 94,100,756 |
| 14 | Vietnam | 86,025,000 | 86,932,500 | 87,840,000 | 88,772,900 | 89,708,900 |
| 15 | Egypt, Arab Rep. | 76,775,023 | 78,075,705 | 79,392,466 | 80,721,874 | 82,056,378 |
| 16 | Germany | 81,902,307 | 81,776,930 | 81,797,673 | 80,425,823 | 80,621,788 |
| 17 | Iran, Islamic Rep. | 73,542,954 | 74,462,314 | 75,424,285 | 76,424,443 | 77,447,168 |
| 18 | Turkey | 71,241,080 | 72,137,546 | 73,058,638 | 73,997,128 | 74,932,641 |
| 19 | Congo, Dem. Rep. | 60,486,276 | 62,191,161 | 63,931,512 | 65,705,093 | 67,513,677 |
| 20 | Thailand | 66,277,335 | 66,402,316 | 66,576,332 | 66,785,001 | 67,010,502 |
| 21 | France | 64,702,921 | 65,023,142 | 65,343,588 | 65,676,758 | 66,028,467 |
| 22 | United Kingdom | 62,276,270 | 62,766,365 | 63,258,918 | 63,695,687 | 64,097,085 |
| 23 | Italy | 59,095,365 | 59,277,417 | 59,379,449 | 59,539,717 | 59,831,093 |
| 24 | Myanmar | 51,540,490 | 51,931,231 | 52,350,763 | 52,797,319 | 53,259,018 |
| 25 | South Africa | 50,222,996 | 50,895,698 | 51,579,599 | 52,274,945 | 52,981,991 |
| 26 | Korea, Rep. | 49,182,038 | 49,410,366 | 49,779,440 | 50,004,441 | 50,219,669 |
| 27 | Tanzania | 43,639,752 | 44,973,330 | 46,354,607 | 47,783,107 | 49,253,126 |
| 28 | Colombia | 45,802,561 | 46,444,798 | 47,078,792 | 47,704,427 | 48,321,405 |
| 29 | Spain | 46,362,946 | 46,576,897 | 46,742,697 | 46,761,264 | 46,647,421 |
| 30 | Ukraine | 46,053,300 | 45,870,700 | 45,706,100 | 45,593,300 | 45,489,600 |

Map of the world population is shown in right part of figure (3). Left map are countries which have been colored based on number of universities in each country.

**Figure 3:**

Map of the world population. Legend: 0 to 50 M to 400 M to 1,336 M (million), 2011: Left map has been created with Google based on our database and right map has been copied from the World Bank.

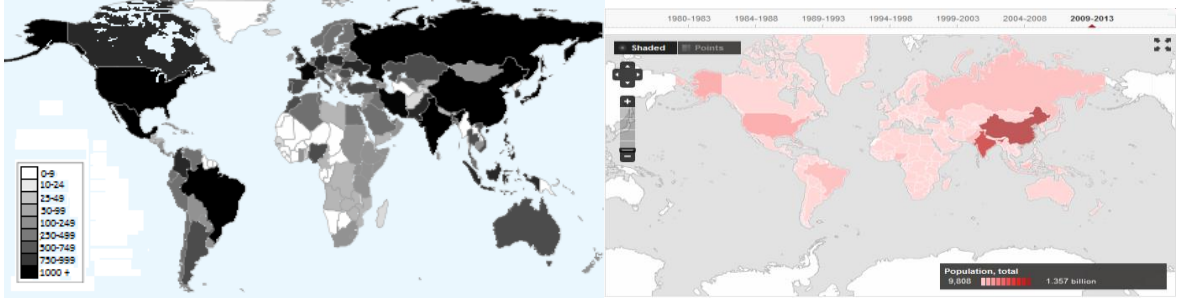

The second candidate is number of universities. Normally, there is a relation between counties' population and number of universities. More population lead to more universities. Each university has its members who only visit its university's site. These members don't visit other sites, then more number of universities don't guarantee to gain better rank. More Population forces to more universities but, doesn't make more visitors for all universities of the country.

Another indicator which has been investigated in this study is language of the countries. Figure (4) shows the percentage of English speakers by countries (right map). This map is more similar to countries which have been colored based on traffic's weight of each country (left map).

**Figure 4:**

Weight of countries' academic traffic (left map) and percentage of English speakers by country (right map. Left map has been created with Google based on our database and right map has been taken from Wikipedia http://en.wikipedia.org/wiki/List_of_countries_by_English-speaking_population)

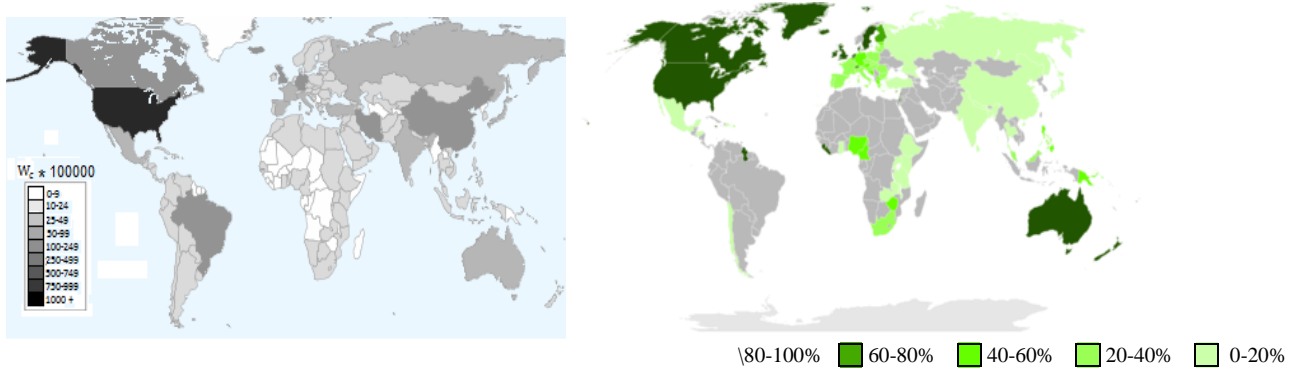

There is a relation between countries' academic traffic and language of sites in each country. Four top countries with higher average of academic traffic in the world speak in English (table (4)).

## 5. Conclusion & future work

Weight of university's traffic could be a good parameter to estimate percentage of universities' real traffic. We can use Weight of country's traffic ($W_c$) for comparing academic traffic of the countries. New indicator percentage of academic traffic ($P_{at}$) is introduced based on weight of university's traffic ($W_u$).

It is recommended to investigate relation between country's academic traffic and speed of internet in the countries. We guess that more internet speed lead to more click. Another good subject to investigate is comparing changes of universities traffic rank in each 2-3 months. Comparing these ranks lead to scholars to investigate effect of registration period of universities and effect of vacations in their Alexa's rank.

**References**

4 International Colleges & Universities. (2012). About us. Retrieved from: http://www.4icu.org/menu/about.htm

Aguillo, I.F., Bar-Ilan, J., Levene, M., & Ortega, J.L. (2010). Comparing university rankings. Scientometrics, 85(1), 243–256.

Aguillo, I. F., Ortega, J. L., & Fernandez, M. (2008). Webometric ranking of world universities: Introduction, methodology, and future developments. Higher Education in Europe, 33(2/3), 234–244.

Alexa the web information company. (2013). About. Retrieved from: http://www.alexa.com/company

Bar-Ilan, J. (2008). Informetrics at the beginning of the 21 stcentury–Areview. Journal of Informetrics, 2(1), 1–52.

Bjǫrneborn, L., & Ingwersen, P. (2004). Toward a basic framework for Webometrics. Journal of the American Society for Information Science and Technology, 55(14), 1216–1227.

Bordons, M., Fernandez, M.T., & Gǒmez, I. (2002). Advantages and limitations in the use of impact factor measures for the assessment of research performance in a peripheral country. Scientometrics, 53(2), 195–206.

Cronin, Blaise. "Bibliometrics and beyond: some thoughts on web-based citation analysis." Journal of Information Science 27.1 (2001): 1-7.

Cronin, Blaise, Elisabeth Davenport, and TAYLOR GRAHAM. "Post-professionalism: Transforming the information heartland." (2013).

Ding, Ying, and Blaise Cronin. "Popular and/or prestigious? Measures of scholarly esteem." Information processing & management 47.1 (2011): 80-96.

.Eccles, C. (2002). The use of university rankings in the United Kingdom. Higher Education in Europe, 27(4), 423–432.

Egghe, L. (2000). New informetric aspects of the Internet: Some reflections, many problems. Journal of Information Science, 26(5), 329–335.

Ghane, M. R., Khosrowjerdi, M., & Azizkhani, Z. (2013). The ranking of Iranian universities based on an improved technique. Malaysian Journal of Library & Information Science, 18(2), 33-45.

Huang, C., & Chang, S. (2009). Commonality of web site visiting among countries. Journal of the American Society for Information Science and Technology, 60(6), 1168–1177.

Huang, M.-H., Lin, C.-S., & Chen, D.-Z. (2011). Counting methods, country rank changes, and counting inflation in the assessment of national research productivity and impact. Journal of the American Society for Information Science and Technology, 62(12), 2427–2436.

Lee, J. Min, J. Oh, A. & Chung, C. (2014). Effective ranking and search techniques for Web resources considering semantic relationships. Information Processing & Management, 50(1), 132-155.

Li, J., Sanderson, M., Willett, P., & Norris, M. (2010). Ranking of library and information science researchers: Comparison of data sources for correlating citation data, and expert judgments. Journal of Informetrics, 4(4), 554–563.

Li, Y., Castellano, C., Radicchi, F., & Ruiz-Castillo, J. (2013). Quantitative evaluation of alternative field normalization procedures. Journal of Informetrics, 7(1), 746-755.

Lin, C. C., & Chien, T. K. (2014). The Effects of Popularity: An Online Store Perspective. *International Journal of Information Science and Management (IJISM)*, 1-11.

Lin, C., Huang, M., & Chen, D. (2013). The influences of counting methods on university rankings based on paper count and citation count. Journal of Informetrics, 7(3), 611–621.

Liu, N. C., & Cheng, Y. (2005). The academic ranking of world universities methodologies and problems. Higher Education in Europe. 30(2), 127–136.

Marginson, S., & van der Wende, M. (2007). To rank or to be ranked: The impact of global rankings in higher education. Journal of Studies in International Education, 11(3/4), 306–329.

Norris, M., &Openheim, C. (2007). Comparing alternatives to the Web of Science for coverage of the social sciences literature. Journal of Informetrics, 1(2), 161–169.

Ortega, J.L. and Aguillo, I.F. (2008b), Visualization of the Nordic academic web: link analysis using social network tools, Information Processing and Management, Vol. 44 No. 4, pp. 1624-33.

Ortega, J.L. and Aguillo, I.F. (2010), Differences between web sessions according to the origin of their visits, Journal of Informetrics. 4 (3), 331-337.

Peeters, H., Verschraegen, G., & Debels, A. (2014). Commensuration and policy comparison: How the use of standardized indicators affects the rankings of pension systems. *Journal of European Social Policy*, *24*(1), 19-38.

Regional information center for science and technology. (2013). About. Retrieved from: http://en.ricest.ac.ir/

Schubert, A., Glonzel, W., & Braun, T. (1987). A new methodology for ranking scientific institutions. Scientometrics, 12, 267-292.

Schubert, A., Glonzel, W., & Braun, T. (1987). A new methodology for ranking scientific institutions. Scientometrics, 12, 267-292.

Thelwall, M., & Vaughan, L. (2004). Webometrics: An introduction to the special issue. Journal of the American Society for Information Science and Technology, 55(14), 1213–1215.

Thelwall, M., Vaughan, L., & Bjӧrneborn, L. (2005). Webometrics. Annual Review of Information Science and Technology, 39, 81–135.

Vaughan, L., & Thelwall, M. (2004). Search engine coverage bias: Evidence and possible causes, Information Processing & Management, 40(4), 693.

Vaughan, L. &Thelwall, M. (2005), A modeling approaches to uncover hyperlink patterns: the case of Canadian universities. Information Processing & Management, 41(2), pp.347-59.

Vaughan, L., Yang, R. (2013). Web traffic and organization performance measures: Relationships and data sources examined. Journal of Informetrics, 7(3), 699–711.

Velikovich, L., Blair-Goldensohn, S., Hannan, K., & McDonald, R. (2010). The viability of web derived polarity lexicons. Proceedings of the 11th Conference of the North American Association for Computational Linguistics, Retrieved June 11, 2011 from: http://www.ryanmcd.com/papers/web_polarity_lexiconsNAACL2010.pdf.

Vieira, E.S., & Gomes, J.A.N.F. (2009). A comparison of Scopus and Web of Science for a typical university. Scientometrics, 81(2), 587–600.

Vinkler, P. (1986). Evaluation of some methods for the relative assessment of scientific publications. Scientometrics, 10, 157-177.

Zitt, M., & Small, H. (2008). Modifying the journal impact factor by fractional citation weighting: The audience factor. Journal of the American Society for Information Science and Technology, 59, 1856-1860.